\documentclass[conference]{IEEEtran}
\IEEEoverridecommandlockouts

\usepackage{algorithmic}
\usepackage{algorithm}
\usepackage{array}
\usepackage{soul}
\usepackage{amsmath,amsfonts,amssymb}
\usepackage[caption=false,font=normalsize,labelfont=sf,textfont=sf]{subfig}
\usepackage{textcomp}
\usepackage{stfloats}
\usepackage{bbm}
\usepackage{url}
\usepackage{verbatim}
\usepackage{graphicx}
\usepackage{cleveref}
\crefname{figure}{Fig.}{Figs.}
\Crefname{figure}{Fig.}{Figs.}
\crefname{equation}{Eq.}{Eqs.}
\Crefname{equation}{Eq.}{Eqs.}
\crefname{appendix}{Appendix}{Appendices}
\Crefname{appendix}{Appendix}{Appendices}
\usepackage{xcolor}
\begin{document}

\title{Efficient, Adaptive Near-Field Beam Training based on Posterior Sampling%
}

\author{
Junchi Liu\textsuperscript{\dag},
Zijun Wang\textsuperscript{\dag},
Rui Zhang\textsuperscript{\dag}\\[0.5ex]
\textsuperscript{\dag}Department of Electrical Engineering, University at Buffalo, SUNY, Buffalo, NY, USA\\
Emails: junchili@buffalo.edu, zwang267@buffalo.edu, rzhang45@buffalo.edu\\
}

\maketitle

\begin{abstract}
This paper proposes a posterior sampling-based beam training framework for near-field communication under multi-path channels. By leveraging Thompson Sampling (TS), the framework adaptively balances exploration and exploitation to maximize the final beamforming gain under a finite pilot overhead. 
To ensure data-efficient learning, we incorporate a structured Gaussian prior in the DFT domain and use an RBF covariance as a tractable
local-correlation prior, motivated by near-field energy leakage, to
promote information sharing among neighboring DFT components.
We develop three TS strategies: codebook-constrained search for rapid stabilization via structural regularization, continuous-space search for high accuracy refinement, and a two-stage hybrid refinement scheme that balances training efficiency and estimation accuracy. Simulation results show that the proposed framework reduces pilot overhead by over 90\% while achieving more than a 2 dB SNR gain over baselines in multipath environments. Furthermore, simulations show that the continuous-space search approaches the full-CSI benchmark as the available pilot numbers become sufficiently large.

\end{abstract}


\section{Introduction}
\IEEEPARstart{T}{he} emerging 6G ecosystem will utilize high-frequency bands such as upper mid band, millimeter-wave (mmWave) and terahertz (THz) bands to meet surging capacity demands \cite{10459211,10438977,wang2023road,singh2025icc}. 
A key enabler for this is extremely large-scale MIMO (XL-MIMO), which employs very large antenna apertures to overcome severe propagation loss. However, large apertures significantly extend the radiative near-field region, where the traditional far-field planar-wave assumption becomes invalid \cite{10942861}. In this region, channel steering vectors are characterized by spherical wavefronts and must be parameterized by both angle and distance for precise beam focusing \cite{zhang2022beam, dai2022channel}.

Near-field beam training typically necessitates exhaustive sweeping over a two-dimensional polar-domain codebook \cite{cui2022farornf}. However, the additional distance dimension leads to a codebook size that scales drastically, incurring prohibitive pilot overhead and training latency \cite{liu2023tutorialreview, cui2023nf6g}. To mitigate this, several efficient strategies have been proposed, often assuming a dominant Line-of-Sight (LoS) path. One representative line of work adopts hierarchical search \cite{10195974,wu2023two} to progressively refine the beam direction and distance without scanning the full 2D codebook. For instance, \cite{wu2023two} proposed a two-stage hierarchical beam training procedure that first searches in the angular domain and then refines the estimate in the polar (angle--range) domain, thereby avoiding exhaustive 2D sweeps. Another line relies solely on far-field DFT beams and infers the range through the intrinsic angle--range coupling. For example, \cite{Joint_Angle_and_Range_Estimation} enable joint angle-and-range estimation via beamwidth information, effectively bypassing 2D polar-domain scanning.
Beyond deterministic search, sequential learning methods based on Bayesian optimization have also been explored to reduce sweeping overhead further. For instance, the Upper Confidence Bound (UCB) framework in \cite{xu2025nearoptimal} models the beam response over near-field codewords using Gaussian process regression; by exploiting the smooth spatial correlation of the channel, such bandit-based methods balance exploration and exploitation to achieve accurate alignment with a small number of pilots \cite{wu2019bandit}.
While effective in reducing overhead, these methods often overlook the impact of multipath propagation. In practice, particularly in indoor, cluttered, or lower-frequency environments such as the upper mid-band, XL-MIMO channels are rarely pure LoS. 
To address the multipath challenge, a multi-beam linear-combination method is proposed in \cite{wang2025multibeam}. By adaptively combining multiple beams based on user-side feedback of the received signal’s amplitude and phase, this approach effectively captures significant propagation paths. However, this approach fundamentally relies on an exhaustive sweep of the entire codebook to ensure reconstruction accuracy, resulting in a pilot overhead comparable to that of traditional methods and limiting its utility in latency-sensitive applications. 

Consequently, how to achieve robust beam alignment under multipath conditions with low pilot overhead is a critical problem. To fill this gap, we propose a posterior sampling beam training framework that employs Thompson Sampling (TS) to efficiently learn the multipath channel and maximize the beamforming gain of the final data beam under a finite pilot overhead. This method can adaptively balance exploration and exploitation.
Motivated by the localized DFT-domain energy spread observed in
near-field channels, we adopt an RBF covariance as a tractable
local correlation prior that promotes information sharing among
neighboring DFT components and accelerates posterior learning.
Furthermore, we develop three TS-based strategies: a codebook-constrained search for rapid initial stabilization, a continuous-space search for high-precision alignment, and a hybrid refinement scheme that adaptively balances training speed and accuracy.

The rest of the paper is organized as follows. \Cref{sec:sys} introduces the system model in the spatial domain and DFT domain. \Cref{sec:ts} proposes three TS-based beam training schemes. \Cref{sec:simulation} analyzes the simulation results and benchmarks our algorithm with the multi-beam combination approach in \cite{wang2025multibeam} and exhaustive near-field codebook search. \Cref{sec:conclusion} concludes the paper.

\section{System Model and Problem formulation}\label{sec:sys}
\begin{figure}[!t]
\captionsetup{justification=justified,singlelinecheck=false} 
\centering
\includegraphics[width=\linewidth]{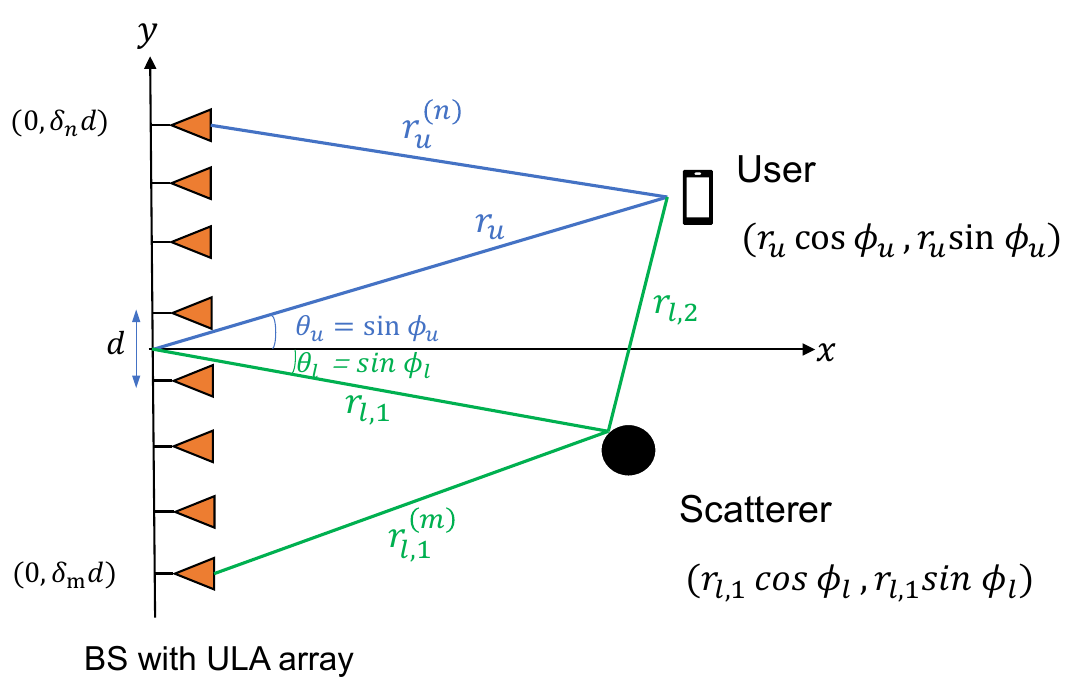}
\caption{Diagram of the near-field communication system featuring a ULA.}
\label{fig:antenna}
\end{figure}

\noindent\hspace*{1em}We consider a system where a base station (BS) equipped with an $N$-element uniform linear array (ULA) serves a single-antenna user equipment (UE), as illustrated in Fig. \ref{fig:antenna}. The boundary between the far-field and near-field regions is typically characterized by the Fresnel distance $R_{\text{Fre}} = \frac{1}{2}\sqrt{\frac{D^3}{\lambda}}$ and the Rayleigh distance $R_{\text{Ray}} = \frac{2D^2}{\lambda}$ \cite{fresnel}. Here, $\lambda$ denotes the carrier wavelength and $D = (N-1)d$ represents the physical aperture of the ULA, where $d$ is the inter-element spacing. 

The system geometry is depicted in \Cref{fig:antenna}. The ULA is aligned along the $y$-axis and centered at the origin $(0,0)$. The $n$-th antenna element is located at $(0, \delta_n d)$, where $\delta_n = \frac{2n-N+1}{2}$ for $n \in \mathcal{N} \triangleq \{0,1,\dots, N-1\}$. Assuming the standard spacing $d = \frac{\lambda}{2}$, the near-field steering vector for a target (UE or scatterer) at distance $r$ and angular direction $\theta$ is defined as:
\begin{equation}
     \mathbf{b}(\theta, r) = \frac{1}{\sqrt{N}}\Biggl[ e^{-j\frac{2\pi (r^{(0)}-r)}{\lambda}},\ldots, e^{-j\frac{2\pi (r^{(N-1)}-r)}{\lambda}} \Biggr]^{T},
     \label{eq:near_field_steering_vector}
\end{equation}
where $r$ is the distance from the UE or scatterer to the array center, and $\theta \triangleq \sin(\phi)$, with $\phi$ being the angle of departure (AoD). The distance between the $n$-th element and the target is given by $r^{(n)} = \sqrt{r^{2} + \delta_n^2 d^{2} - 2r\delta_n d \theta}$.

We consider a multipath channel comprising $L$ total paths. For a user located at $(\theta_u,r_u)$, the Line-of-Sight (LoS) component is modeled as
\begin{equation}
   \mathbf{h}_{\text{LoS}}
   = \sqrt{N}\gamma_u
   e^{-j\frac{2\pi r_u}{\lambda}}
   \mathbf{b}(\theta_u,r_u),
   \label{eq:LoS channel_model}
\end{equation}
where $\gamma_u = \frac{\lambda}{4\pi r_u}$
denotes the path loss gain. The Non-Line-of-Sight (NLoS) component involves $L-1$ scatterers and is expressed as
{\setlength{\abovedisplayskip}{3pt}
\setlength{\belowdisplayskip}{3pt}
\setlength{\abovedisplayshortskip}{3pt}
\setlength{\belowdisplayshortskip}{3pt}
\begin{equation}
   \mathbf{h}_{\text{NLoS}}
   =
   \sqrt{N}
   \sum_{l=1}^{L-1}
   \gamma_l
   e^{-j\frac{2\pi(r_{l,1}+r_{l,2})}{\lambda}}
   \mathbf{b}(\theta_l,r_{l,1}),
   \label{eq:NLoS channel_model}
\end{equation}}
where the $l$-th scatterer is positioned at $(\theta_l, r_{l,1})$ and its distance to the user is $r_{l,2}=\sqrt{r_{l,1}^{2} + r_u^{2} - 2r_u r_{l,1}\cos(\phi_u - \phi_l)}$ via the law of cosines. The complex gain $\gamma_l=\frac{\lambda p_l}{4\pi(r_{l,1}+r_{l,2})}$ accounts for path loss and the random reflection coefficient $p_l \sim \mathcal{CN}(0,1)$. 
Consequently, the total multipath channel model is given by $\mathbf{h}=\mathbf{h}_{\text{LoS}}+\mathbf{h}_{\text{NLoS}}.$

Hence, the received signal $y$ at a user located at $(\theta_u, r_u)$ 
is formulated as:
\begin{equation}
  y(\mathbf{w})=\mathbf{w}^{H}\mathbf{h} x+n,  
  \label{eq:received signal model}
\end{equation}
where $\mathbf{w} \in \mathbb{C}^{N \times 1}$ denotes the beamforming vector and $n \sim \mathcal{CN}(0, \sigma^2)$ represents the complex additive white Gaussian noise (AWGN). Without loss of generality, the reference pilot signal is set to $x = 1$.

For each channel realization, we define the SNR as the
average per-antenna channel power normalized by the noise variance:
\begin{equation}
\rho \triangleq \frac{\|\mathbf h\|_2^2}{N\sigma^2},
\qquad
\rho_{\mathrm{dB}}
=10\log_{10}\rho.
\label{eq:snr_definition}
\end{equation}
Accordingly, for a prescribed $\rho_{\rm dB}$, the noise variance used
in the simulations is $\sigma^2
=
\|\mathbf{h}\|_2^2 / \left(N10^{\rho_{\rm dB}/10}\right).$
This SNR is independent of the selected training beam.  The
instantaneous receive SNR produced by a unit-norm beam $\mathbf w$ is
$|\mathbf w^H\mathbf h|^2/\sigma^2$.

To represent the channel in the angular domain and expose its beamspace structure, we introduce a unitary DFT matrix $\mathbf{F} \in \mathbb{C}^{N \times N}$. We define the DFT-domain channel coefficient vector $\mathbf{g}$, which is related to the spatial-domain channel vector $\mathbf{h}$ by:
{\setlength{\abovedisplayskip}{-2pt}
\setlength{\belowdisplayskip}{3pt}
\setlength{\abovedisplayshortskip}{-2pt}
\setlength{\belowdisplayshortskip}{3pt}
\begin{equation}\label{eq:beamspace_def}
    \mathbf{h}=\mathbf{F}^H\mathbf{g}.
\end{equation}}
The unitary DFT matrix is defined as $\mathbf{F} = [\mathbf{a}(\varphi_0), \mathbf{a}(\varphi_1), \dots, \mathbf{a}(\varphi_{N-1})]$, where the directional-cosine grid points $\varphi_n$ are uniformly sampled such that $\varphi_n = \frac{2n-N+1}{N}$ for $n \in \mathcal{N}$. The $n$-th column of $\mathbf{F}$ is given by:
{\setlength{\abovedisplayskip}{3pt}
\setlength{\belowdisplayskip}{0pt}
\setlength{\abovedisplayshortskip}{3pt}
\setlength{\belowdisplayshortskip}{0pt}
\begin{align}
 &\mathbf{a}(\varphi_n) \nonumber\\
 &\triangleq \frac{1}{\sqrt{N}} 
\left[ 
e^{-j\pi \left(-\frac{N-1}{2}\right)\varphi_n},
\cdots,e^{-j\pi \left(\frac{N-1}{2}\right)\varphi_n}
\right]^T,\forall n \in \mathcal{N}.   
 \label{eq:DFT codeword}
\end{align}}

Substituting \Cref{eq:beamspace_def} into \Cref{eq:received signal model} yields a linear Gaussian observation model for the DFT-domain channel $\mathbf{g}$: $y = \mathbf{v}^H \mathbf{g} + n,
\mathbf{v} \triangleq \mathbf{F}\mathbf{w}.$


For a unit-norm beamformer $\mathbf w$, we define the normalized
beamforming gain as
{\setlength{\abovedisplayskip}{3pt}
\setlength{\belowdisplayskip}{3pt}
\setlength{\abovedisplayshortskip}{3pt}
\setlength{\belowdisplayshortskip}{3pt}
\begin{equation}
G(\mathbf w)
\triangleq |\mathbf w^H\mathbf h|^2 / \|\mathbf h\|_2^2,
\qquad 0\leq G(\mathbf w)\leq 1.
\label{eq:normalized_gain}
\end{equation}}
With full CSI, the maximum gain is attained by
$\mathbf{w}_{\mathrm{CSI}}
=\mathbf{h}/\|\mathbf{h}\|_2$.
The BS sequentially transmits at most $T$ pilot beams. At slot
$t$, the training beam $\mathbf{w}_t$ is selected based on the
previous observation history $\mathcal H_{t-1} = \{(\mathbf w_s,y_s)\}_{s=1}^{t-1}.$
Let $\Pi_T$ denote the class of adaptive beam-training policies
that use no more than $T$ pilot transmissions and produce a
final data beam $\mathbf{w}_{\mathrm{data}}^{\pi}$. Under the
hard maximum pilot number $T$, the primary beam-training
objective is
{\setlength{\abovedisplayskip}{3pt}
\setlength{\belowdisplayskip}{3pt}
\setlength{\abovedisplayshortskip}{3pt}
\setlength{\belowdisplayshortskip}{3pt}
\begin{equation}
    \max_{\pi\in\Pi_T}
    \;
    \mathbb{E}\!\left[
        G\!\left(\mathbf{w}_{\mathrm{data}}^{\pi}\right)
    \right].
    \label{eq:beam_training_objective}
\end{equation}}
The practical termination rule introduced in Section~III may
stop training before $T$. We therefore report both the final
achievable rate and the average realized pilot overhead.



\section{Proposed TS-based Beam Training}\label{sec:ts}
To facilitate Thompson Sampling (TS), we model the DFT-domain channel
coefficient vector as a proper complex Gaussian random vector:
{\setlength{\abovedisplayskip}{3pt}
\setlength{\belowdisplayskip}{3pt}
\setlength{\abovedisplayshortskip}{3pt}
\setlength{\belowdisplayshortskip}{3pt}
\begin{equation}\label{eq:prior_g}
\mathbf{g}\sim\mathcal{CN}(\mathbf{m}_0,\mathbf{D}_0).
\end{equation}}

To state precisely which structure is encoded by \(\mathbf D_0\), we extend
the discrete DFT coefficients to the continuous directional-cosine variable
\(\varphi\):
\begin{equation}\label{eq:continuous_beamspace}
g(\varphi)
\triangleq
\frac{1}{\sqrt N}\sum_{n=0}^{N-1}h_n
 e^{-j\pi\delta_n\varphi},
\qquad g_k=g(\varphi_k).
\end{equation}
where $h_n$ denotes the $n$-th entry of $\mathbf{h}$. The function \(g(\varphi)\) is a finite trigonometric polynomial. Hence, for
any \(\varphi,\varphi'\in[-1,1]\),
{\setlength{\abovedisplayskip}{3pt}
\setlength{\belowdisplayskip}{3pt}
\setlength{\abovedisplayshortskip}{3pt}
\setlength{\belowdisplayshortskip}{3pt}
\begin{equation}\label{eq:beamspace_lipschitz}
\left|g(\varphi)-g(\varphi')\right|
\leq
\pi\sqrt{\frac{N^2-1}{12}}\,
\|\mathbf h\|_2\,|\varphi-\varphi'|,
\end{equation}}
where we used
\(\sum_{n=0}^{N-1}\delta_n^2=N(N^2-1)/12\). Thus, the complex beamspace
response varies continuously along the angular coordinate. An exactly
on-grid far-field plane wave is a special case whose DFT representation is
concentrated in one bin. By contrast, the nonlinear phase profile of a
near-field spherical wave generally produces a response over several
adjacent DFT bins \cite{cui2022near}. 
The continuity in \Cref{eq:beamspace_lipschitz}, together with the near-field multi-bin
response, motivates imposing a local smoothness bias on the complex
samples $g(\varphi_k)$. Accordingly, we adopt the following
one-parameter RBF covariance as a tractable surrogate prior:
{\setlength{\abovedisplayskip}{3pt}
\setlength{\belowdisplayskip}{3pt}
\setlength{\abovedisplayshortskip}{3pt}
\setlength{\belowdisplayshortskip}{3pt}
\begin{equation}
    [\mathbf{D}_0]_{i,j}
    =
    \sigma_0 \left( \frac{\lambda}{4 \pi \sqrt{r_{min} r_{max}}} \right)^2
    \exp\left(
        -\frac{(\varphi_i-\varphi_j)^2}{2\ell^2}
    \right),
\end{equation}}
where $\sigma_0$ is a hyperparameter controlling the prior variance and $\ell$ is the RBF length scale. $r_{min}$ and $r_{max}$ are the minimum and maximum range of the searching area. The RBF covariance
is a parsimonious surrogate prior that imposes local mean-square smoothness in the DFT directional-cosine domain.

Let $\bar{\mathbf{g}}
    \triangleq
    \mathbf{g}-\mathbf{m}_0$
denote the centered DFT-domain channel. For the proper complex prior
in \Cref{eq:prior_g}, $\operatorname{Re}\{\bar{\mathbf{g}}\}$ and
$\operatorname{Im}\{\bar{\mathbf{g}}\}$ are independent real Gaussian
vectors with covariance $\mathbf{D}_0/2$. Moreover,
{\setlength{\abovedisplayskip}{3pt}
\setlength{\belowdisplayskip}{3pt}
\setlength{\abovedisplayshortskip}{3pt}
\setlength{\belowdisplayshortskip}{3pt}
\begin{equation} \label{eq:local evidence}
\begin{aligned}
    \mathbb{E}\left[
        \left|\bar{g}_i-\bar{g}_j\right|^2
    \right]
    =
    2A_0
    \left[
        1-
        \exp\left(
            -\frac{(\varphi_i-\varphi_j)^2}{2\ell^2}
        \right)
    \right].
\end{aligned}
\end{equation}}
where $A_0 = \sigma_0 \left( \frac{\lambda}{4 \pi \sqrt{r_{min} r_{max}}} \right)^2$. Thus, under the adopted prior, nearby DFT coefficients are assigned smaller mean-square differences than distant coefficients. 
\Cref{eq:local evidence} therefore provides a probabilistic interpretation of the local regularization imposed by the RBF kernel.

A small $\ell$ makes $\mathbf{D}_0$ close to a scaled identity matrix
and therefore provides limited information sharing among neighboring
components. In contrast, an excessively large $\ell$ imposes strong
prior coupling over multiple directional-cosine bins, which may
oversmooth spatially separated multipath components and may lead to
premature posterior concentration. Therefore, $\ell$ controls a
tradeoff between data-efficient local information sharing and
preservation of angular resolution. In all simulations,
$\sigma_0=1$ and $\mathbf{m}_0 = \mathbf{0}$ are fixed across the compared TS schemes.

The posterior distribution of $\mathbf{g}$, given the history, is modeled as a complex Gaussian distribution:
{\setlength{\abovedisplayskip}{3pt}
\setlength{\belowdisplayskip}{3pt}
\setlength{\abovedisplayshortskip}{3pt}
\setlength{\belowdisplayshortskip}{3pt}
\begin{equation}\label{eq:g-distribution}
\mathbf{g}\mid \mathcal{H}_{t-1} \sim \mathcal{CN}(\mathbf{m}_{t-1},\mathbf{D}_{t-1}).
\end{equation}}
Here, $\mathbf{m}_{t-1}$ is the posterior mean in the DFT domain, and the corresponding spatial-domain estimate is $\hat{\mathbf{h}}_{t-1} \triangleq \mathbf{F}^H \mathbf{m}_{t-1}$. At slot $t$, we draw $\tilde{\mathbf{g}}_t \sim \mathcal{CN}(\mathbf{m}_{t-1},\mathbf{D}_{t-1})$ and map it to the spatial domain as $\tilde{\mathbf{h}}_t=\mathbf F^H\tilde{\mathbf{g}}_t$. The sampled channel $\tilde{\mathbf{h}}_t$ is used, either directly or through the codebook, to select the training beam $\mathbf{w}_t^\star$ for slot $t$.
After transmitting $\mathbf{w}_t^\star$ and observing the resulting signal $y_t$, the belief (mean and covariance of the posterior distribution) is updated iteratively. This framework transforms the beam training task into a sequential search: in each slot, we sample from the current belief to choose a beam and subsequently refine that belief based on the new observation.

We leverage TS to adaptively balance exploration and exploitation by drawing samples from the posterior distribution in Eq. \eqref{eq:g-distribution}.
In the Gaussian posterior, $\mathbf{D}_{t-1}$ quantifies the uncertainty about the unknown channel $\mathbf g$. Its diagonal entries are the marginal posterior variances of the corresponding DFT-domain coefficients. Unlike other algorithms, such as UCB, TS inherently balances exploration and exploitation via randomness drawn from the posterior. When the posterior uncertainty is large, the Thompson samples $\widetilde{\mathbf{h}}_t$ exhibit greater variability, promoting exploration. As the posterior concentrates, the samples become increasingly concentrated around the posterior-mean estimate $\hat{\mathbf{h}}_t$ and the algorithm naturally favors exploitation.

As additional pilot observations are collected, the posterior covariance
matrix $\mathbf D_t$ quantifies the posterior uncertainty about the unknown DFT-domain channel $\mathbf{g}$ at time $t$.  In
particular, $\operatorname{tr}(\mathbf D_t)$ equals the sum of the marginal posterior variances over all DFT-domain channel coefficients. Therefore, we determine the termination criterion from the reduction of this total uncertainty over a window of
$W$ pilot transmissions:
\begin{equation}
\eta_t \triangleq
\frac{\operatorname{tr}(\mathbf D_{t-W})-
      \operatorname{tr}(\mathbf D_t)}
     {\operatorname{tr}(\mathbf D_{t-W})}
\le \tau, \qquad t\ge W,
\label{eq:trace_stopping}
\end{equation}
where $\tau$ is the threshold associated with the active probing scheme
or stage. The normalization expresses the uncertainty decrease relative to the
start of the current window; the prior scale is held fixed in all comparisons.  For Scheme III, the trace window
is reset when Stage I switches to Stage II, and the Stage II threshold is
then applied to the subsequent continuous-space refinement.

We then consider three TS-based beam-training schemes for selecting the transmit beam during training process.

\begin{itemize}
[%
    \setlength{\topsep}{0pt}%
    \setlength{\partopsep}{0pt}%
    \setlength{\parsep}{0pt}%
    \setlength{\itemsep}{0pt}%
    \setlength{\parskip}{0pt}%
]
    \item \textbf{Scheme I (Codebook-constrained TS).}
    The action set of beam sweeping in this case is restricted to the near-field codebook $\mathcal{W}$ defined in \cite{cui2022farornf}. This near-field codebook is constructed in the polar domain using uniform angular sampling and non-uniform distance sampling. For each sweeping time slot, TS selects the codeword that maximizes its inner product with the sampled channel from the posterior distribution: 
    {\setlength{\abovedisplayskip}{3pt}
    \setlength{\belowdisplayskip}{3pt}
    \setlength{\abovedisplayshortskip}{3pt}
    \setlength{\belowdisplayshortskip}{3pt}
    \begin{equation}
        \mathbf{w}_t^\star = \arg\max_{\mathbf{w}\in\mathcal{W}} \, |\mathbf{\widetilde{h}}_t^H \mathbf{w}|,
        \label{eq:scheme1_select}
    \end{equation}}
    Once the stopping criterion in \Cref{eq:trace_stopping} is satisfied at time slot $T'$, the data beam is constructed from the latest posterior mean $\mathbf{m}_{T'}$ as, 
    \begin{equation}\label{eq:ts_data} \mathbf{w}_{\mathrm{data}} = \hat{\mathbf{h}}_{T'} / \|\hat{\mathbf{h}}_{T'}\|_2 = \mathbf{F}^{H}\mathbf{m}_{T'} / \|\mathbf{F}^{H}\mathbf{m}_{T'}\|_2. \end{equation}
    The near-field codebook provides an intrinsic regularization effect. By constraining the search space, Scheme I limits large deviations of the probing beam under high initial channel uncertainty, thereby promoting rapid stabilization and reducing the required training overhead, particularly in low-SNR regimes.    
    \item \textbf{Scheme II (Continuous Space TS).}
    Unlike scheme I, the action set of beam sweeping is the full unit sphere \(\mathcal{W}_{\rm cont}=\{\mathbf{w}\in\mathbb{C}^N:\|\mathbf{w}\|_2=1\}\). TS directly uses the normalized sampled channel for the beam training:
    {\setlength{\abovedisplayskip}{3pt}
    \setlength{\belowdisplayskip}{3pt}
    \setlength{\abovedisplayshortskip}{1pt}
    \setlength{\belowdisplayshortskip}{3pt}
    \begin{equation}\label{eq:ts_scheme2_select} \mathbf{w}_t^\star = \widetilde{\mathbf{h}}_t / \|\widetilde{\mathbf{h}}_t\|_2. \end{equation}
    }        
    Once the stopping criterion is satisfied, the data beam is constructed in the same way as \Cref{eq:ts_data}. 
    By removing the codebook constraint while beam sweeping, Scheme II can exploit a richer set of probing directions and approach the full-CSI benchmark more closely when sufficiently many pilots are available, although it may require substantially more training.
    \item \textbf{Scheme III (Hybrid refinement TS).}
    To consolidate the advantages of both approaches, we propose a Hybrid TS scheme that operates in two consecutive stages. 
   In Stage 1, it follows Scheme I and performs TS-based beam training over the codebook $\mathcal{W}$ according to Eq. \eqref{eq:scheme1_select}. This exploits the regularization of the codebook to quickly stabilize the search. Once the Stage-I trace criterion in \Cref{eq:trace_stopping} is satisfied, the action set switches to the continuous space $\mathcal W_{\rm cont}$.
    At the switching slot, the trace window is reset and the Stage-II threshold $\tau_{\mathrm{III},2}$ is activated. The training beam is then selected according to \Cref{eq:ts_scheme2_select} for continuous-space refinement. Once the Stage-II stopping criterion is satisfied at time slot $T'$, the data beam is formed using \Cref{eq:ts_data}. This hybrid approach uses the codebook to warm-start the estimation process, providing rapid initial stabilization before transitioning to continuous-space refinement for improved beamforming accuracy.
\end{itemize}

For all three schemes, if the termination criterion is not satisfied within the maximum $T$ pilots, the final data beam is constructed from the latest posterior mean as: $\mathbf{w}_{\mathrm{data}}
=
\hat{\mathbf{h}}_{T} / \|\hat{\mathbf{h}}_{T}\|_2
=
\mathbf{F}^{H}\mathbf{m}_{T} / \left\|\mathbf{F}^{H}\mathbf{m}_{T}\right\|_2.$
All three schemes share the same Bayesian update rules. After transmitting $\mathbf{w}_t^\star$ and receiving the observation $y_t$, we set
$\boldsymbol{\mathbf{v}}_t=\mathbf{F}\mathbf{w}_t^\star$ and 
update the posterior mean and covariance $(\mathbf{m}_{t-1},\mathbf{D}_{t-1})\mapsto(\mathbf{m}_{t},\mathbf{D}_{t})$ as follows:
{\setlength{\abovedisplayskip}{3pt}
\setlength{\belowdisplayskip}{3pt}
\setlength{\abovedisplayshortskip}{3pt}
\setlength{\belowdisplayshortskip}{3pt}
\begin{align}
\alpha_t &\triangleq \boldsymbol{\mathbf{v}}_t^H \mathbf{D}_{t-1}\boldsymbol{\mathbf{v}}_t + \sigma^2,
\label{eq:alpha_update}\\
\mathbf{k}_t &\triangleq \mathbf{D}_{t-1}\mathbf{v}_t / \alpha_t,
\label{eq:gain_update}\\
\mathbf{m}_{t} &= \mathbf{m}_{t-1} + \mathbf{k}_t\Big(y_t-\boldsymbol{\mathbf{v}}_t^H\mathbf{m}_{t-1}\Big),
\label{eq:mean_update}\\
\mathbf{D}_{t} &= \mathbf{D}_{t-1} - \mathbf{k}_t \boldsymbol{\mathbf{v}}_t^H \mathbf{D}_{t-1}.
\label{eq:cov_update}
\end{align}}
The updated spatial-domain channel estimate is consequently
$\hat{\mathbf{h}}_{t} \triangleq \mathbf{F}^H \mathbf{m}_{t}$. The local
information-sharing mechanism follows directly from \Cref{eq:gain_update}.
For example, under a canonical DFT probe
$\boldsymbol{\mathbf v}_t=\mathbf e_i$, where $\mathbf e_i$ denotes the
$i$-th canonical basis vector, the $j$-th entry of the Bayesian gain
is
$[\mathbf k_t]_j=[\mathbf D_{t-1}]_{j,i}/
([\mathbf D_{t-1}]_{i,i}+\sigma^2)$; hence an observation centered on bin
$i$ updates neighboring bins in proportion to their current posterior
covariance. Since $y_t$ and $\mathbf m_t$ are complex-valued, the recursion
in \Cref{eq:mean_update} jointly updates the in-phase and quadrature
components, equivalently the amplitudes and phases, of the DFT-domain
coefficients.
\section{Simulation Results and Analysis}\label{sec:simulation}

\noindent\hspace*{1em}To evaluate the proposed schemes, we consider a BS equipped with $N=512$ antenna elements operating at a carrier frequency of $100$ GHz. The channel model comprises $L=4$ paths, consisting of one LoS and three NLoS components. The user and scatterers are randomly distributed within a distance range of $[9, 380]$ m and an angular range of $[-\pi/3, \pi/3]$. We employ the near-field polar codebook $\mathcal{W}$ from \cite{cui2022farornf} with an expansion factor $\beta = 1.1$, resulting in a total of $2560$ codewords. For the TS-based beam training, the covariance-trace window is set to $W=10$ pilot transmissions. The thresholds are
$\tau_{\mathrm I}=10^{-5}$ for Scheme I,
$\tau_{\mathrm{II}}=10^{-2}$ for Scheme II, and $(\tau_{\mathrm{III},1},\tau_{\mathrm{III},2})
=(10^{-5},10^{-2})$ for the two stages of Scheme III. 
These thresholds are selected based on the probing space. For codebook-constrained probing, a smaller threshold is used because the trace reduction may become small while the probing beams remain concentrated around a suboptimal codeword. A loose threshold may therefore satisfy \Cref{eq:trace_stopping} prematurely. For continuous-space probing, a larger threshold is used because the richer action space can sustain gradual posterior uncertainty reduction for a longer period. This allows timely stopping once further refinement provides limited benefit.
Unless otherwise specified, the reference SNR is defined by
\Cref{eq:snr_definition}. All results are averaged over $N_{\rm iter}=1000$ independent Monte Carlo trials to ensure statistical robustness. The performance is quantified by the achievable rate, defined as: $R = \log_2 \left(1 + \frac{\left| \mathbf{w}_{\mathrm{data}}^H \mathbf{h} \right |^2}{\sigma^2}  \right )$.
In this section, we evaluate the performance of the three proposed TS-based schemes alongside the multi-beam linear combination approach introduced in \cite{wang2025multibeam}. The latter performs an exhaustive sweep with a far-field DFT codebook and then forms a data beam by linearly combining the measured DFT beams. For benchmarking, we also include (i) full-CSI case, which is the upper bound under the unit-norm beam constraint, and (ii) conventional exhaustive search over the near-field (NF) codebook $\mathcal{W}$, whose beam is restricted to a single codeword.

\begin{figure}
    \centering
    \includegraphics[width=\linewidth]{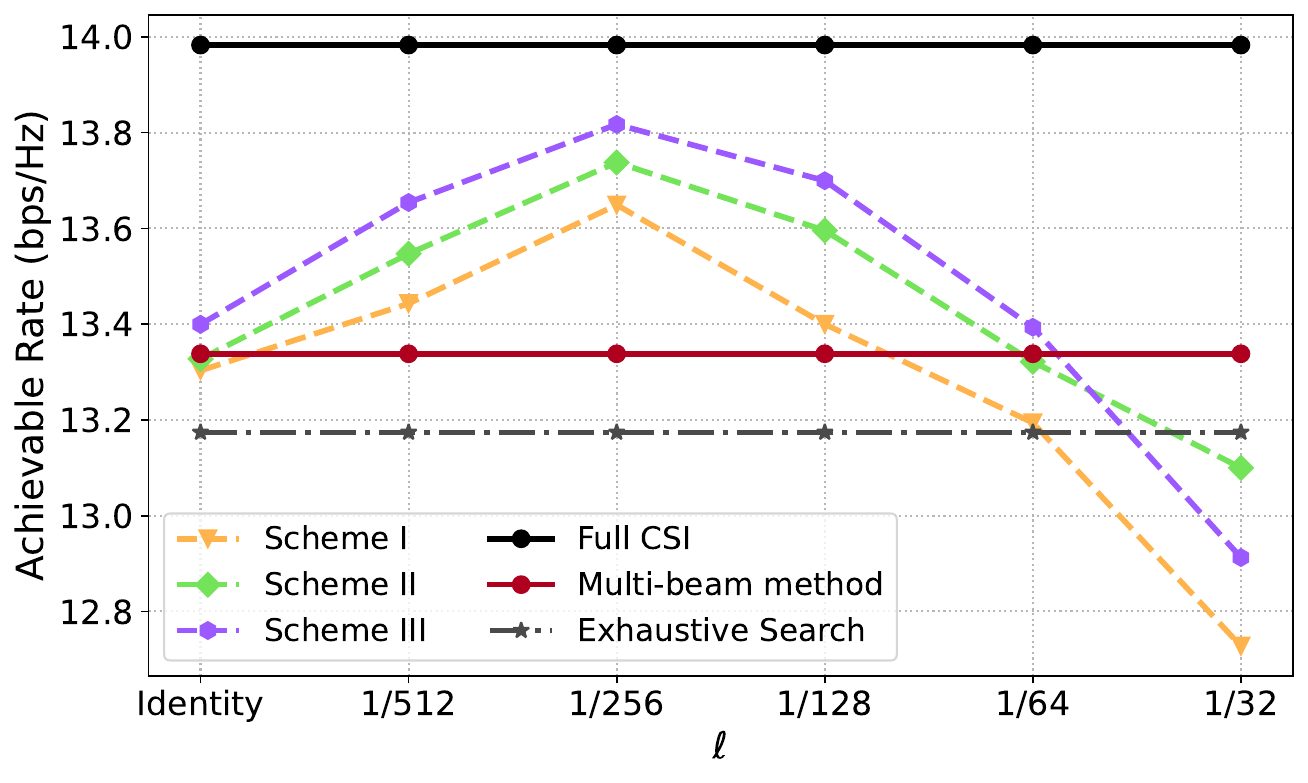}
    \caption{Achievable rate versus the RBF length scale $\ell$ at SNR =
$15$ dB.}
    \label{fig:ell_sensitivity}
\end{figure}

\begin{figure}[t]
    \centering
    \includegraphics[width=\linewidth, height=5.5cm]{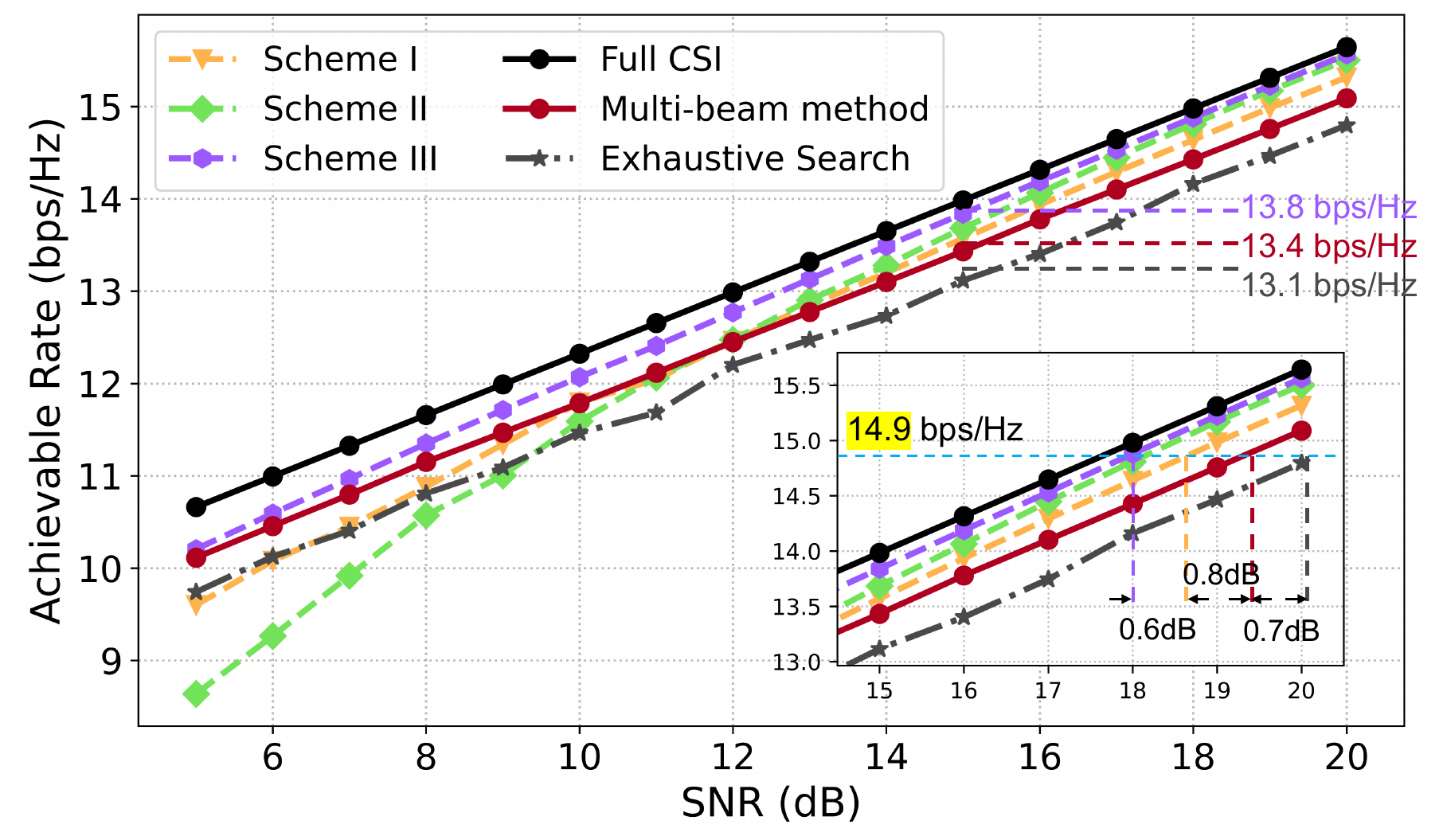}
    \vspace{-5mm}
    \caption{Achievable rate performance vs. SNR of various schemes under finite maximum pilot overhead constraints.} 
    \label{fig:rate-snr}
\end{figure}

\Cref{fig:ell_sensitivity} evaluates the sensitivity of the
proposed schemes to the RBF length scale $\ell$ at $\rho_{\mathrm{dB}}=15$ dB, with
$\mathbf D_0= A_0 \mathbf I$ included as an independent-prior baseline.
Introducing a finite local coupling scale improves the achievable rate of
all three schemes under the finite pilots. The rates first increase
with $\ell$ and attain their maxima at
$\ell=1/256$, where Schemes I, II, and III achieve
approximately $13.6$, $13.7$, and $13.8$ bps/Hz, respectively. When $\ell$
is increased further, the achievable rates decrease, indicating that a
broader prior coupling neighborhood is less effective for the considered
multipath channel. We therefore use $\ell=1/256$ in the
remaining simulations.

\begin{figure}
    \centering
    \includegraphics[width=\linewidth]{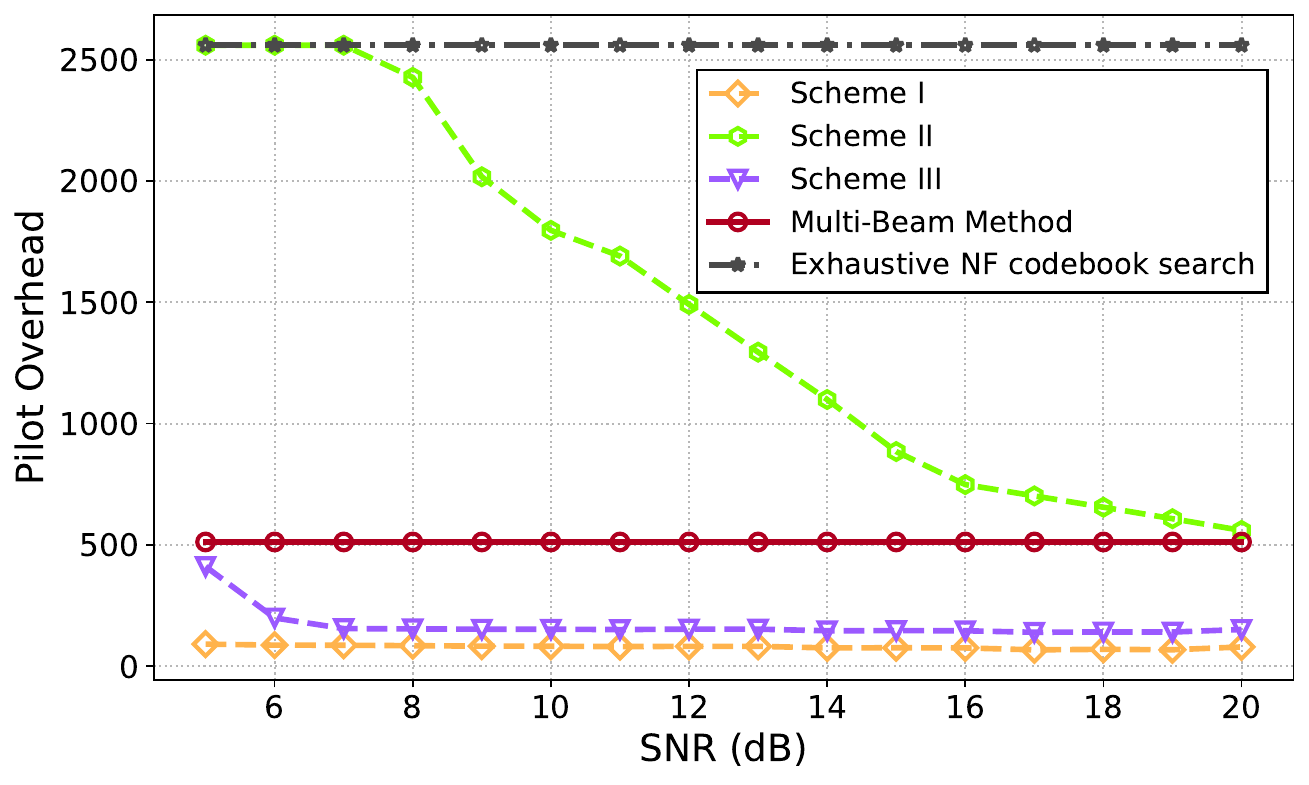}
    \caption{Beam Training Pilot Overhead versus SNR of various schemes.}
    \label{fig:overhead}
\end{figure}

Fig. \ref{fig:rate-snr} illustrates the achievable rate versus SNR for the evaluated schemes, while Fig. \ref{fig:overhead} presents the corresponding training pilot overhead for SNR levels ranging from $5$ to $20$~dB. The pilot overhead refers to the number of pilot symbols transmitted during beam training and the maximum pilot number is $T = 2560$ (matching the codebook size). Both the proposed TS-based frameworks and the multi-beam method significantly outperform conventional exhaustive search due to their ability to capture multiple paths. Among the proposed approaches, the hybrid refinement TS (Scheme III) exhibits the most robust performance, maintaining the narrowest gap to the full-CSI upper bound across the entire SNR regime. Specifically, at an SNR of $15$~dB, Scheme III achieves a rate of $13.8$~bps/Hz, surpassing the multi-beam method ($13.4$~bps/Hz) and exhaustive search ($13.1$~bps/Hz). This spectral efficiency gain is achieved with an average pilot overhead of only $146.5$, which is substantially lower than the $512$ and $2560$ pilots required by the multi-beam and exhaustive baselines, respectively. Furthermore, to attain a target achievable rate of $14.9$~bps/Hz, Scheme III offers SNR gains of approximately $0.6$~dB, $1.4$~dB, and over $2$~dB relative to Scheme I, the multi-beam baseline, and exhaustive NF codebook search, respectively.

Conversely, Scheme II degrades markedly in the low-SNR regime. Its unconstrained probing is more sensitive to noise and often does not satisfy the stopping criterion within $T$, as illustrated in Fig. \ref{fig:overhead}. However, as the SNR increases, Scheme II more frequently satisfies the termination criterion within $T$ and eliminates the grid quantization errors inherent in discrete codebooks, eventually surpassing the codebook-constrained Scheme I. By combining codebook-constrained probing for rapid stabilization with continuous-space refinement, Scheme III achieves the best finite-pilots performance among the proposed schemes over the evaluated SNR range. It reduces pilot overhead by approximately $90\%$ compared to a full NF codebook sweep. Notably, while the multi-beam benchmark necessitates a constant overhead of $512$ pilots for its initial DFT-domain exhaustive search, the proposed TS framework dynamically reduces its sweeping duration, highlighting its superior pilot efficiency for latency-sensitive near-field communications.

\begin{figure}
    \centering
    \includegraphics[width=\linewidth]{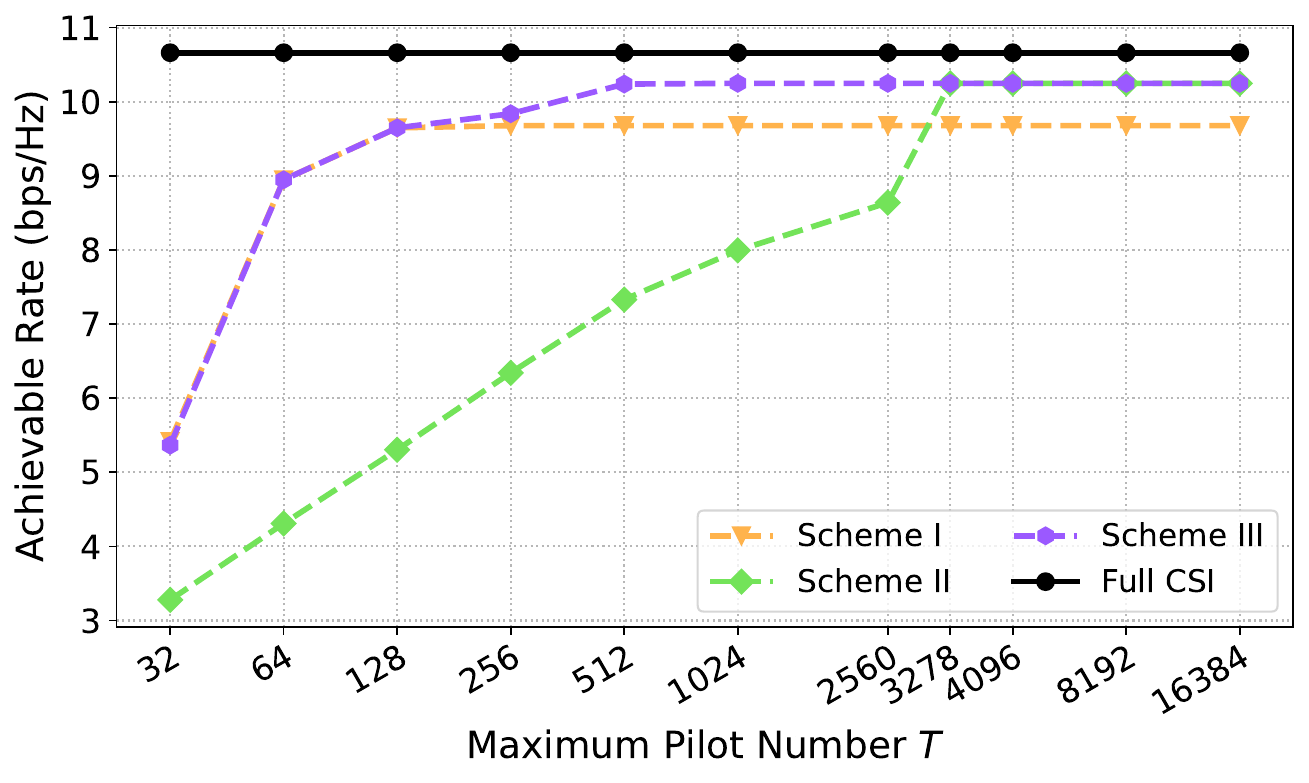}
    \caption{Achievable rate versus the maximum pilot number for three proposed TS schemes.}
    \label{fig:rate_snr_budget}
\end{figure}

Next, we vary the maximum pilot number $T$ to investigate the performance of the three proposed TS schemes in the low-SNR regime. The SNR is fixed at 5 dB.

The effect of the maximum pilot number on the achievable rate is illustrated in \Cref{fig:rate_snr_budget}.
At $T=128$, Scheme~I requires only $90.8$ pilot transmissions on average before
termination and achieves an achievable rate of approximately $9.6$~bps/Hz.
Its performance remains nearly unchanged as $T$ further increases. This behavior
indicates that allowing additional pilots provides little further improvement once
the stopping criterion is triggered, while the codebook-constrained action space limits the refinement of off-grid multipath components.
For Scheme~III, when $T=512$, the average number of pilot transmissions before
termination is $410.7$, and the achievable rate increases to approximately
$10.2$~bps/Hz. The additional pilot transmissions are used for continuous-space
refinement after the codebook-based initialization, thereby mitigating the resolution
limitation of Scheme~I.
In contrast, Scheme~II requires approximately $3278.4$ pilot transmissions on
average before the \Cref{eq:trace_stopping} is satisfied while $T$ is over $4096$. Consequently, its performance
improves more gradually as $T$ increases and approaches the full-CSI benchmark only
when a substantially larger number of pilot transmissions is allowed.
These results show that Scheme~I terminates after relatively few pilot transmissions
but is ultimately limited by the codebook resolution, whereas Scheme~II can achieve
higher performance for sufficiently large $T$ at the cost of substantially greater
training overhead. Scheme~III balances well between training
efficiency and beamforming accuracy by combining codebook-based training with
continuous-space refinement.

The results further reveal three distinct operating regimes as $T$ increases.
For small $T$, Scheme~I benefits from codebook-assisted probing and reaches a
performance plateau relatively quickly, whereas Scheme~II devotes more pilot
transmissions to exploration and therefore exhibits lower achievable rates.
For intermediate $T$, Scheme~III preserves the efficient initialization of Scheme~I
while further improving the achievable rate through continuous-space refinement.
For sufficiently large $T$, Scheme~II continues to improve and eventually approaches
the full-CSI benchmark.

However, Scheme~II should be viewed as an idealized hardware benchmark because its
unit-sphere action set permits arbitrary complex antenna weights. Implementing these
weights may require fully digital beamforming or a sufficiently flexible hybrid/analog
architecture, which means extremely high cost on RF chains.
Therefore, although Scheme~II approaches full-CSI benchmark when sufficiently many
pilot transmissions and flexible hardware resources are available, Scheme~III offers
a more practical tradeoff between training efficiency and beamforming accuracy.
\section{Conclusion}\label{sec:conclusion}
\noindent\hspace*{1em} This paper presented a posterior sampling framework for efficient near-field beam training in XL-MIMO systems under multipath propagation. By leveraging Thompson Sampling in the DFT domain, we used a proper complex Gaussian prior with an RBF covariance to impose local mean-square smoothness on the complex DFT-domain response. Built upon Bayesian recursion, three strategies were developed: codebook-constrained search for rapid stabilization, continuous-space search for high-precision alignment, and a hybrid scheme that balances both. Simulation results demonstrate that the proposed framework reduces pilot overhead by approximately one order of magnitude compared with the exhaustive sweeping baseline, while remaining robust in multipath environments. Besides, it indicate that, continuous-space posterior sampling also approaches the full-CSI benchmark as the maximum pilot number increases at low SNR.

\bibliographystyle{IEEEtran}
\bibliography{ref}

\end{document}